# Title: Longitudinal Assessment of Seasonal Impacts and Depression Associations on Circadian Rhythm Using Multimodal Wearable Sensing

Yuezhou Zhang[1], PhD; Amos A Folarin[1], PhD; Shaoxiong Sun[1], PhD; Nicholas Cummins[1], PhD; Yatharth Ranjan[1], MSc; Zulqarnain Rashid[1], PhD; Callum Stewart[1], PhD; Pauline Conde[1], BSc; Heet Sankesara[1], BSc; Petroula Laiou[1], PhD; Faith Matcham[2], PhD; Katie M White[2], BSc; Carolin Oetzmann[2], MSc; Femke Lamers[3], PhD; Sara Siddi[4], PhD; Sara Simblett[5], PhD; Srinivasan Vairavan[6], PhD; Inez Myin-Germeys[7], PhD; David C. Mohr[8], PhD; Til Wykes[5], PhD; Josep Maria Haro[4], PhD; Peter Annas[9], PhD; Brenda WJH Penninx[3], PhD; Vaibhav A Narayan[6], PhD; Matthew Hotopf[2], PhD; Richard JB Dobson[1], PhD; RADAR-CNS consortium[10]

[1]Department of Biostatistics & Health Informatics, Institute of Psychiatry, Psychology and Neuroscience, King's College London, London, United Kingdom
[2]Department of Psychological Medicine, Institute of Psychiatry, Psychology and Neuroscience, King's College London, London, United Kingdom
[3]Department of Psychiatry, Amsterdam UMC location Vrije Universiteit, Amsterdam, Netherlands
[4]Centro de Investigación Biomédica en Red de Salud Mental, Madrid, Spain
[5]Department of Psychology, Institute of Psychiatry, Psychology and Neuroscience, King's College London, London, United Kingdom
[6]Janssen Research and Development LLC, Titusville, NJ, United States
[7]Department of Neurosciences, Center for Contextual Psychiatry, Katholieke Universiteit Leuven, Leuven, Belgium
[8]Department of Preventive Medicine, Northwestern University, Chicago, IL, United States
[9]H Lundbeck A/S, Copenhagen, Denmark
[10]www.radar-cns.org

**Corresponding Authors:**
Yuezhou Zhang Address: Department of Biostatistics & Health Informatics, Institute of Psychiatry, Psychology and Neuroscience, King's College London, De Crespigny Park, Denmark Hill, London, SE5 8AF. Email: yuezhou.zhang@kcl.ac.uk. Phone: 44 75 7985 6617
Richard Dobson Address: Department of Biostatistics & Health Informatics, Institute of Psychiatry, Psychology and Neuroscience, King's College London, De Crespigny Park, Denmark Hill, London, SE5 8AF. Email: richard.j.dobson@kcl.ac.uk. Phone: 44 20 7848 0473

**ABSTRACT**

**Objective:** This study aimed to explore the associations between depression severity and wearable-measured circadian rhythms, accounting for seasonal impacts and quantifying seasonal changes in circadian rhythms.

**Materials and Methods:** Data used in this study came from a large longitudinal mobile health study. Participants' depression severity (measured biweekly using the 8-item Patient Health Questionnaire [PHQ-8]) and behaviors (monitored by Fitbit) were tracked for up to two years. Twelve features were extracted from Fitbit recordings to approximate circadian rhythms. Three nested linear mixed-effects models were employed for each feature: (1) incorporating the PHQ-8 score as an independent variable; (2) adding the season variable; and (3) adding an interaction term between season and the PHQ-8 score.

**Results:** This study analyzed 10,018 PHQ-8 records with Fitbit data from 543 participants. Upon adjusting for seasonal effects, higher PHQ-8 scores were associated with reduced activity, irregular behaviors, and delayed rhythms. Notably, the negative association with daily step counts was stronger in summer and spring than in winter, and the positive association with the onset of the most active continuous 10-hour period was significant only during summer. Furthermore, participants had shorter and later sleep, more activity, and delayed circadian rhythms in summer compared to winter.

**Discussion and Conclusions:** Our findings underscore the significant seasonal impacts on human circadian rhythms and their associations with depression and indicate that wearable-measured circadian rhythms have the potential to be the digital biomarkers of depression.

# INTRODUCTION

**Background and significance**

Circadian rhythms are approximately 24-hour endogenous oscillations, controlled by the master clock in the suprachiasmatic nucleus (SCN) of the hypothalamus, that regulate many aspects of human behavior and physiology, such as sleep-wake cycles, hormone secretion, and body temperature.[1-3] Disturbances in circadian rhythms have been associated with an increased risk of both physical and mental diseases, highlighting the critical role of well-regulated circadian rhythms for overall health and well-being.[4-7]

Circadian rhythm disturbances have been strongly linked to depression.[7, 8] As a globally prevalent mental disorder, depression can lead to a range of negative outcomes, including diminished quality of life, disability, premature mortality, and suicide.[9-14] However, current depression diagnosis relies heavily on subjective measures like questionnaires and interviews,[15, 16] resulting in the underdiagnosis and undertreatment of individuals with depression.[17, 18] Tracking human circadian rhythms is a potential objective method for early-stage depression identification. However, the gold standard for circadian rhythm estimation involves tracking melatonin in bodily fluids under a controlled light condition,[19, 20] which is expensive, labor-intensive, and impractical for large cohort studies and long-term monitoring in real-world settings.[21, 22] There is a critical need for easy-to-use approaches to measure individual circadian rhythms in the real world.

Wearable devices provide a convenient and cost-effective way to continuously monitor individuals' behaviors and physiological signals in real-world settings.[23] Previous mobile health (mHealth) studies have explored the approximation of human circadian rhythms through wearable-measured patterns, including sleep-wake cycles, rest-activity patterns, and circadian rhythms in heart rate (HR), and revealed their associations with depression severity.[24-30] However, seasonal effects were not fully considered in these mHealth studies, potentially due to their short study durations. Seasonal changes in sunlight and temperature are crucial environmental zeitgebers for the internal circadian clock, impacting human circadian rhythms.[31-33] Prior research has reported significant seasonal effects on sleep patterns and activity levels.[34-37] Disregarding seasonal effects may introduce bias into the associations between depression and wearable-measured circadian rhythms in real-world settings. Therefore, there is a need to explore the seasonal effects on the wearable-measured circadian rhythms and their associations with depression in a large longitudinal dataset.

**Objective**

The primary aim of this study was to explore the associations between depression severity and wearable-measured circadian rhythms, accounting for seasonal effects and investigating potential variations across seasons. Our secondary aim was to quantify the seasonal changes in wearable-measured circadian rhythms within a European mHealth study for depression.[38]

**MATERIALS AND METHODS**

**Participants and settings**

The data analyzed used in this study came from the Remote Assessment of Disease and Relapse Major Depressive Disorder (RADAR-MDD) research program, which aimed to investigate the utility of remote technologies for monitoring depression and understanding factors that could help predict relapse in MDD.[38] A total of 623 participants were recruited from three study sites across three European countries (United Kingdom, Spain, and the Netherlands) and followed for up to 2 years.[39] Recruitment was conducted between November 2017 and June 2020, with data collection concluding in April 2021.[39] The RADAR-MDD program employed the RADAR-base open-source platform to concurrently gather both active (e.g., questionnaires) and passive (e.g., wearable) data.[40]

**Patient involvement**

The RADAR-MDD protocol was co-developed with a patient advisory board (PAB) who shared their opinions on several user-facing aspects of the study including the choice and frequency of survey measures, the usability of the study app, participant-facing documents, selection of optimal participation incentives, selection, and deployment of wearable device as well as the data analysis plan.

**Measures**

**Depression Symptom Severity.** Participants' depression symptom severity was measured using the 8-item Patient Health Questionnaire (PHQ-8),[41] conducted via mobile phones every two weeks. The PHQ-8 comprises eight questions and the total score of PHQ-8 ranges from 0 to 24, indicating increasing severity.[41]

**Fitbit Data.** Participants were asked to wear a Fitbit Charge 2/3 wrist-worn device during the follow-up period. Participants' sleep, step count, and HR were continuously measured and recorded. <u>Sleep data:</u> Fitbit provided sleep labels ("awake", "light sleep", "deep sleep", and "rapid eye movement") along with the corresponding local clock times every 30 seconds. <u>Step data:</u>

Participants' accumulated steps were counted every minute. HR data: Fitbit provided an estimate of HR every 5 seconds, utilizing an embedded photoplethysmography sensor. However, technical issues resulted in the absence of some sample points. To obtain the robust HR trend and align with step data, we calculated the average heart rate over one minute.

**Season.** The seasonal division used in this study was based on EU astronomical seasons: spring (March 20—June 20), summer (June 21— September 22), autumn (September 23—December 20), and winter (December 21 — March 19).

**Feature extraction of wearable circadian rhythms**

**A 14-day PHQ-8 interval.** To link human circadian rhythms with depression severity, we extracted circadian rhythm features from each 14-day PHQ-8 interval — 14 days of Fitbit recordings before a completed PHQ-8, as the PHQ-8 is used to measure depressive symptom severity for the past 2 weeks.[41] Figure 1 shows an example of a participant's processed HR, step, and sleep data in a 14-day PHQ-8 interval.

**Data Inclusion Criteria.** Some Fitbit data were missing in our dataset for several reasons, including device damage, low battery level, and not being worn. Building on insights from our prior research,[42] which discussed the sufficient days for stable feature calculation, we focused on 14-day PHQ-8 intervals with at least 8 days having more than 80% of step and HR data and a sleep recording in the present study. Considering the potential impact of daylight saving time on individuals' behaviors,[43] we excluded the first 14-day PHQ-8 interval after the time switching.[35] We then extracted a total of 12 features for reflecting the circadian rhythms from sleep, step count, and HR data.

**Sleep-Wake Rhythms.** Fitbit shows promise in identifying sleep-wake status.[44, 45] Therefore, to reflect the sleep-wake rhythms, we computed four features: (1) *Sleep Duration*—the mean total sleep time, (2) *Sleep Variability*—the standard deviation of total sleep time, (3) *Sleep Onset*—the mean clock time of falling asleep, and (4) *Sleep Offset*— the mean clock time of wake-up.[46]

**Rest-Activity Rhythms.** We extracted five nonparametric features from the Fitbit step count recordings to characterize the stability, fragmentation, timing, and mean activity level of participants' rest-activity rhythms, utilizing the R package "nparACT".[47, 48] These features include: (1) intradaily variability of steps (*Step IV*) — quantifying the fragmentation in rest-activity cycle, (2) interdaily stability of steps (*Step IS*) — quantifying the stability of rest-activity patterns over a 14-day PHQ-8 interval, (3) *L5 Onset*—representing the onset of least active continuous 5-hour period, (4) *M10 Onset*—representing the onset of the most active continuous 10-hour period, and (5) *Daily Step*—representing the mean of daily total steps in a 14-day PHQ-8 interval.[47, 48]

**Circadian Rhythm in HR.** For estimating circadian rhythms in HR, we employed Cosinor analysis—fitting a cosine wave to time series behavioral data through least-squares regression, which has been widely used in previous mHealth studies.[49-51] Using the R package "cosinor", we performed the Cosinor analysis on Fitbit HR data of each 14-day PHQ-8 interval and extracted the following parameters: (1) *HR MESOR*—the midline estimating statistic of the fitted cosine wave for HR, (2) *HR Amplitude*—the difference between the peak value and MESOR of the fitted cosine wave for HR, and (3) *HR Acrophase*—the timing of the HR peak.[49-51]

**Covariates**

In accordance with findings from previous studies,[52-54] we considered several covariates that could potentially influence participants' circadian rhythms, including age, gender, and employment status. Since the COVID-19 pandemic and relevant restrictions had some significant impacts on individuals' behavior,[55] we introduced a covariate "lockdown" to indicate the presence of a national lockdown. Furthermore, as the experience of seasons can be different across countries, the study site was also considered as a covariate. These covariates were considered in our statistical analysis.

**Statistical analysis**

Given the longitudinal nature of our dataset, i.e., each participant had repeated measurements, we utilized the linear mixed-effects model[56] with a participant-specific random intercept in this study, implemented using the R package "lmerTest". To investigate whether disregarding seasonal effects biases the associations between depression severity and circadian rhythms, we established and compared the following three models for each of circadian rhythm features.

**Models 1:** a linear mixed-effect model was established to regress each circadian rhythm feature with only the PHQ-8 score as the independent variable. **Model 2:** season was included as an independent variable in addition to the PHQ-8 score, considering seasonal effects on the circadian rhythms feature. **Model 3:** To further explore potential variations in the association between depression severity and circadian rhythms across seasons, an interaction term between the PHQ-8 score and season was added to the main effects model (Model 2). All models were adjusted by covariates: age, gender, study site, lockdown, and employment status. The equations of these 3 models are outlined as follows:

**Model 1:** $Circadian\ rhythm = \beta_1 PHQ8 + COVs$

**Model 2:** $Circadian\ rhythm = \beta_1 PHQ8 + \beta_2 season + COVs$

**Model 3:** $Circadian\ rhythm = \beta_1 PHQ8 + \beta_2 season + \beta_3 PHQ8 \times season + COVs$

where COVs represents all covariates mentioned above and $Circadian\ rhythm$ is one of the wearable-measured circadian rhythm features.

Likelihood ratio tests were then performed to examine whether including more variables (season and the interaction term) can significantly improve the fitting of the regression model. The Benjamini-Hochberg method was used for the correction of multiple comparisons.[57] Despite the inclusion of lockdown as a covariate, we also repeated our analysis on the subset of data before the COVID-19 pandemic (before 1st March 2020).

## RESULTS

**Data summary**

According to our data inclusion criteria (see MATERIALS AND METHODS), a total of 10,018 PHQ-8 records with corresponding Fitbit data from 543 participants (each participant had 16 recordings on average) were analyzed in this study. Table 1 summarizes participant demographics and PHQ-8 records. The selected cohort has a median age (IQR) of 48 (32, 58) years, with a majority of female (76.2%, N=414) and 230 (42.4%) employed participants. Approximately one-quarter of PHQ-8 records were collected from each season (winter: 26.9%, spring: 23.6%, summer:25.5%, and autumn: 23.9%). Figure 2 visualizes the variations of wearable-measured circadian rhythm features across a year.

**Table 1.** A summary of sociodemographics of participants and PHQ-8 questionnaires in the present study.

| Characteristic | Statistics |
|---|---|
| Participant, n | 543 |
| Age, median (IQR) | 48.00 [32.00, 58.00] |
| Female, n (%) | 414 (76.2%) |
| Employed, n (%) | 230 (42.4%) |
| PHQ-8 records, n | 10,018 |
| PHQ-8 records per participant, median (IQR) | 16.00 [7.00, 27.50] |
| Records in Winter, n (%) | 2699 (26.9%) |
| Records in Spring, n (%) | 2363 (23.6%) |
| Records in Summer, n (%) | 2559 (25.5%) |
| Records in Autumn, n (%) | 2397 (23.9%) |
| Records before COVID-19 pandemic, n (%) | 4202 (41.9%) |

**Associations between depression severity and circadian rhythm features**

According to the likelihood ratio tests, we found that incorporating seasonal effects (Model 2 or Model 3) significantly improved the goodness-of-fit for all 12 circadian rhythm features (Table 2). Notably, the inclusion of the interaction term between PHQ-8 and season significantly refined the regression for *Daily Step*, *Step IV, Step IS*, *M10 Onset*, *HR MESOR*, and *HR Amplitude*. Specifically, the negative association between the PHQ-8 score and *Daily Step* exhibited greater strength in spring ($\beta_3$ of PHQ8×Spring = -31.51, $P < 0.01$) and summer ($\beta_3$ of PHQ8×Summer = -42.61, $P < 0.001$) compared to winter. *M10 Onset* demonstrated a significant positive association with the PHQ-8 score only in summer ($\beta_3$ of PHQ8×Summer = 1.06, $P < 0.01$), with no significant association observed in other seasons. The effect sizes of *Step IV, Step IS*, *HR MESOR*, and *HR Amplitude* are small, according to their practical meanings. Notably, the association between *HR Acrophase* and the PHQ-8 score was weaker in Model 1 ($\beta_1 = 0.51$, $P < 0.05$) than in Model 2 ($\beta_1 = 0.71$, $P < 0.01$). Furthermore, the associations between the remaining features and PHQ-8 were similar across models with and without seasonal effects. Specifically, the PHQ-8 score was positively associated with *Sleep Duration* ($\beta_1 = 0.46$, $P < 0.001$), *Sleep Onset* ($\beta_1 = 0.55$, $P < 0.01$), *Sleep Offset* ($\beta_1 = 1.12$, $P < 0.001$), *Sleep Variability* ($\beta_1 = 0.96$, $P < 0.001$), and *L5 Onset* ($\beta_1 = 0.46$, $P < 0.001$).

**Table 2.** Associations between the PHQ-8 score and wearable-measured circadian rhythms in the entire dataset, estimated by 3 nested linear mixed-effects models.

| Feature | Model 1[a] | Model 2[b] | Model 3[c] | | | | LR test |
|---|---|---|---|---|---|---|---|
| | PHQ8 $\beta_1$(SE) | PHQ8 $\beta_1$(SE) | PHQ8 $\beta_1$(SE) | PHQ8×Spring $\beta_3$(SE) | PHQ8×Summer $\beta_3$(SE) | PHQ8×Autumn $\beta_3$(SE) | |
| Sleep Duration | 0.51(0.14) *** | 0.46(0.14) *** | 0.49(0.18) ** | -0.20(0.22) | -0.04(0.22) | 0.11(0.21) | Model 2 |
| Sleep Onset | 0.52(0.17) ** | 0.55(0.17) ** | 0.79(0.22) *** | -0.09(0.26) | -0.41(0.26) | -0.53(0.26) * | Model 2 |
| Sleep Offset | 1.13(0.15) *** | 1.12(0.15) *** | 1.01(0.20) *** | 0.27(0.24) | 0.04(0.23) | 0.15(0.23) | Model 2 |
| Sleep Variability | 0.97(0.14) *** | 0.96(0.14) *** | 1.23(0.19) *** | -0.42(0.22) | -0.57(0.22) ** | -0.14(0.22) | Model 2 |
| Mean Step | -94.71(6.43) *** | -93.61(6.41) *** | -73.21(8.55) *** | -31.51(10.01) ** | -42.61(9.95) *** | -12.66(9.87) | Model 3 |
| Step IV | 0.001(0.0006) | 0.0007(0.0006) | -0.001(0.0008) | 0.002(0.001) * | 0.004(0.001) *** | 0.002(0.001) | Model 3 |
| Step IS | -0.001(0.0002) *** | -0.001(0.0002) *** | -0.001(0.0003) ** | -0.001(0.0004) * | -0.001(0.0004) | -0.0001(0.0004) | Model 3 |
| L5 Onset | 0.43 (0.21) * | 0.43(0.21) * | 0.51(0.29) | 0.31(0.35) | -0.54(0.35) | -0.11 (0.34) | Model 2 |
| M10 Onset | 0.66(0.25) ** | 0.73(0.25) ** | 0.26(0.34) | 0.72(0.41) | 1.06 (0.40) ** | 0.21(0.40) | Model 3 |
| HR MESOR | -0.03(0.01) *** | -0.03 (0.01) *** | -0.03(0.01) * | 0.01(0.01) | -0.01(0.01) | -0.03(0.01) | Model 3 |
| HR Amplitude | -0.04(0.01) *** | -0.04 (0.01) *** | -0.04 (0.01) *** | -0.02(0.01) * | 0.01(0.01) | 0.002(0.008) | Model 3 |
| HR Acrophase | 0.51(0.24) * | 0.71(0.22) ** | 0.50(0.30) | 0.30(0.35) | 0.52(0.35) | 0.08(0.35) | Model 2 |

Coefficient estimates (standard error [SE]) of PHQ-8 and interaction term between PHQ-8 and season are shown in this table. Likelihood ratio tests were performed to compare the fitness of these 3 models and the model with best fitness are shown in this table. * $P < 0.05$, ** $P < 0.01$, *** $P < 0.001$.

[a]Model 1: $Circadian\ rhythm = \beta_1 PHQ8 + COVs$

[b]Model 2: $Circadian\ rhythm = \beta_1 PHQ8 + \beta_2 season + COVs$

[c]Model 3: $Circadian\ rhythm = \beta_1 PHQ8 + \beta_2 season + \beta_3 PHQ8 \times season + COVs$, where COVs represents covariates mentioned in Methods.

**Seasonal changes in circadian rhythm features**

Table 3 presents the coefficients of seasons of Model 2, indicating significant variations in various circadian rhythm features across seasons. Specifically, we found substantial seasonal differences in *HR Acrophase*, with spring delaying it by 43.4 minutes ($P < 0.001$), summer by 67.9 minutes ($P < 0.001$), and autumn by 24.1 minutes ($P < 0.001$) compared to winter (the reference season).

Regarding rest-activity rhythms, compared to winter, summer was associated with 394.5 more daily steps ($P < 0.001$), 0.03 lower *Step IV* ($P < 0.001$), and 20.5 minutes later *M10 Onset* ($P < 0.001$); autumn was associated with 186.2 more daily steps ($P < 0.01$) and 0.02 lower *Step IV* ($P < 0.001$); and spring was associated with 11.9 minutes later *M10 Onset* ($P < 0.001$).

For sleep-wake rhythms, compared to winter, we found that: (i) *Sleep Duration* was 6.5 minutes shorter in spring ($P < 0.001$), 16.6 minutes shorter in summer ($P < 0.001$), and 6.1 minutes shorter in autumn ($P < 0.001$); (ii) *Sleep Onset* was 4.6 minutes earlier in autumn ($P < 0.01$), 3.7 minutes later in spring ($P < 0.05$), and 7.2 minutes later summer ($P < 0.001$); (iii) *Sleep Offset* was 6.9 minutes earlier in summer ($P < 0.001$), and 8.3 minutes earlier in autumn ($P < 0.001$); (iv) *Sleep Variability* was 8.5 minutes lower in spring ($P < 0.001$), 5.3 minutes lower in summer ($P < 0.001$), and 2.8 minutes lower in autumn ($P < 0.05$).

**Table 3.** Seasonal changes in circadian rhythm features in the entire dataset and pre-COVID subset, estimated by linear mixed-effects models (Model 2).

| Feature | Entire Dataset | | | Pre-COVID Subset | | |
|---|---|---|---|---|---|---|
| | Spring $\beta_2$(SE) | Summer $\beta_2$(SE) | Autumn $\beta_2$(SE) | Spring $\beta_2$(SE) | Summer $\beta_2$(SE) | Autumn $\beta_2$(SE) |
| Sleep Duration | -6.48(1.32) *** | -16.63(1.31) *** | -6.13(1.31) *** | -3.06(2.14) | -15.35(2.01) *** | -5.75(1.84) ** |
| Sleep Onset | 3.68(1.61) * | 7.16(1.59) *** | -4.64(1.60) ** | -4.65(2.92) | 4.77(2.75) | -5.92(2.52) * |
| Sleep Offset | -0.25(1.44) | -6.91(1.43) *** | -8.32(1.43) *** | -8.51(2.49) *** | -9.76(2.34) *** | -9.36(2.14) *** |
| Sleep Variability | -8.48(1.36) *** | -5.29 (1.35) *** | -2.82(1.35) * | -5.64(2.19) * | -6.48(2.06) ** | -2.73(1.89) |
| Mean Step | -19.21(61.28) | 394.46(60.81) *** | 186.16(60.75) ** | 456.92(96.28) *** | 622.87(90.68) *** | 336.36(83.01) *** |
| Step IV | 0.001(0.006) | -0.026(0.006) *** | -0.021(0.006) *** | -0.008(0.009) | 0.001(0.009) | -0.003(0.008) |
| Step IS | 0.002(0.002) | -0.006(0.002) * | 0.004(0.002) | -0.007(0.003) | -0.004(0.003) | 0.001(0.003) |
| L5 Onset | -1.10(2.13) | -3.84(2.12) | -3.98(2.12) | -13.88(3.82) *** | -8.67(3.59) * | -4.03(3.31) |
| M10 Onset | 11.94(2.45) *** | 20.51(2.43) *** | 1.61(2.44) | 3.94(4.22) | 16.71(3.97) *** | 1.97(3.65) |
| HR MESOR | -0.47(0.09) *** | -0.01(0.09) | 0.04(0.09) | -0.26(0.14) | -0.09(0.14) | 0.01(0.12) |
| HR Amplitude | 0.19(0.05) *** | 0.92(0.05) *** | 0.39(0.05) *** | 0.18(0.08) * | 0.78(0.08) *** | 0.30(0.08) *** |
| HR Acrophase | 43.35(2.15) *** | 67.94(2.13) *** | 24.09(2.13) *** | 46.48(3.09) *** | 70.67(2.91) *** | 24.15(2.67) *** |

Note, coefficient estimates (standard error) of season variable are displayed in this table; winter is the reference season. * $P < 0.05$, ** $P < 0.01$, *** $P < 0.001$.

**Pre-COVID subset analysis**

In examining the pre-COVID subset, our results revealed a consistency in the direction and significance of associations between depression severity and circadian rhythm features compared to the entire dataset. Notably, sleep-related features, *L5 Onset, M10 Onset,* and *HR Acrophase* displayed stronger associations with the PHQ-8 score in the pre-COVID subset compared to the entire dataset (Table 4).

The seasonal changes in most circadian rhythm features (excluding *Daily Step*) were similar between the entire dataset and the pre-COVID subset (Table 3). Noteworthy is the relatively larger seasonal changes in *Daily Step* in the pre-COVID subset compared to the entire dataset. In the pre-COVID subset, compared to winter, participants exhibited 456.9, 622.9, and 336.4 more daily steps in spring ($P < 0.001$), summer ($P < 0.001$), and autumn ($P < 0.001$), respectively.

**Table 4.** Associations between the PHQ-8 score and wearable-measured circadian rhythms in the pre-COVID subset, estimated by 3 nested linear mixed-effects models.

| Feature | Model 1[a] | Model 2[b] | Model 3[c] | | | | LR test |
|---|---|---|---|---|---|---|---|
| | PHQ8 $\beta_1$(SE) | PHQ8 $\beta_1$(SE) | PHQ8 $\beta_1$(SE) | PHQ8×Spring $\beta_3$(SE) | PHQ8×Summer $\beta_3$(SE) | PHQ8×Autumn $\beta_3$(SE) | |
| Sleep Duration | 1.08(0.20) *** | 1.01(0.20) *** | 0.99(0.26) *** | -0.01(0.34) | -0.05 (0.31) | 0.11(0.29) | Model 2 |
| Sleep Onset | 0.66(0.28) * | 0.70(0.28) * | 1.05(0.36) ** | -0.22(0.46) | -0.50(0.43) | -0.67(0.39) | Model 2 |
| Sleep Offset | 1.53(0.24) *** | 1.50(0.23) *** | 1.57(0.31) *** | -0.22(0.40) | -0.22(0.36) | 0.05(0.34) | Model 2 |
| Sleep Variability | 1.23(0.19) *** | 1.21(0.19) *** | 1.63(0.26) *** | -0.74(0.35) * | -0.88(0.32) ** | -0.35(0.29) | Model 2 |
| Mean Step | -88.11(9.43) *** | -84.92 (9.37) *** | -71.04(12.01) *** | -20.43(15.21) | -48.62 (13.99) *** | 1.43(12.91) | Model 3 |
| Step IV | -0.0003(0.0009) | -0.0003(0.0009) | -0.002(0.001) | 0.003(0.002) * | 0.003(0.001) * | 0.0007(0.001) | Model 3 |
| Step IS | -0.0004(0.0004) | -0.0004(0.0004) | -0.0002(0.0005) | -0.0003(0.0006) | -0.0006(0.0006) | -0.0003(0.0005) | Model 3 |
| L5 Onset | 0.78(0.33) * | 0.75(0.33) * | 0.78(0.45) | 0.43(0.61) | -0.17(0.56) | -0.20 (0.52) | Model 2 |
| M10 Onset | 0.77(0.38) * | 0.85(0.38) * | 0.37(0.50) | 1.18(0.67) | 1.25(0.62) * | 0.03 (0.57) | Model 3 |
| HR MESOR | -0.05(0.01) *** | -0.06(0.01) *** | -0.08(0.02) *** | 0.01(0.02) | 0.04(0.02) * | 0.04(0.02) * | Model 3 |
| HR Amplitude | -0.04(0.01) *** | -0.04(0.01) *** | -0.05(0.01) *** | -0.01(0.01) | 0.02(0.01) | 0.02 (0.01) | Model 3 |
| HR Acrophase | 0.68(0.32) * | 1.03(0.30) *** | 0.75(0.38) | 1.17(0.49) * | 0.25(0.45) | 0.16 (0.42) | Model 2 |

Coefficient estimates (standard error [SE]) of PHQ-8 and interaction term between PHQ-8 and season are shown in this table. Likelihood ratio tests were performed to compare the fitness of these 3 models and the model with best fitness are shown in this table. * $P < 0.05$, ** $P < 0.01$, *** $P < 0.001$.

[a]Model 1: $Circadian\ rhythm = \beta_1 PHQ8 + COVs$

[b]Model 2: $Circadian\ rhythm = \beta_1 PHQ8 + \beta_2 season + COVs$

[c]Model 3: $Circadian\ rhythm = \beta_1 PHQ8 + \beta_2 season + \beta_3 PHQ8 \times season + COVs$, where COVs represents covariates mentioned in Methods.

## DISCUSSION

We report findings regarding the seasonal impacts on circadian rhythms and their associations with depression from a large European longitudinal mHealth study for depression. One of our key findings is that the associations between depression severity and certain wearable-measured circadian rhythms differ across seasons. Specifically, we observed a stronger negative association between depression and daily step count in summer and spring compared to winter. Additionally, depression severity exhibited a significant and positive association with the onset of the most active continuous 10-hour period exclusively in summer. The potential reason for the above findings is that the climate in summer is more suitable for outdoor activities,[58] so the associations between rest-activity rhythms and depression could be better reflected. We also found the association between the peak hour of HR and depression severity was underestimated without considering the seasonal effects. Although the associations of the remaining features with depression were similar across models with and without seasonal effects, the likelihood ratio tests illustrated that incorporating seasonal effects significantly improved the model fitness. These findings highlight the critical importance of including seasonal effects in longitudinal mHealth studies.

Upon adjusting for seasonal effects, we identified that higher depression severity was significantly associated with lower activity levels (*Daily Step*), irregular activities (*Sleep Variability* and *Step IS*), and later timings of rhythms (*Sleep Offset*, *M10 Onset,* and *HR Acrophase*). These relationships can be supported by prior studies. The linkage between lower physical activity levels and higher depression severity was reported in both survey-based and mHealth studies.[59, 60] The preventive and therapeutic potential of physical activity against depression has also been reported.[61, 62] Prior research has extensively documented the connections between depression and irregular daily behaviors, such as increased sleep variability,[46, 63-65] lower interdaily stability of rest-activity rhythm,[66] and irregular patterns in Bluetooth[67] and GPS[68] data. The delayed timing of circadian rhythms has been found to be associated with higher depression severity in multiple data streams, including sleep onset and offset,[46, 63-65] *M10 Onset*,[51] and *HR Acrophase*.[69] These consistent results indicate the close associations between circadian rhythms and depression severity.

This study revealed significant changes in circadian rhythms across seasons, with the most differences observed between summer and winter. Participants in our cohort exhibited shorter and later sleep, increased daily step counts, lower intradaily variability of steps, and delayed timings of circadian rhythms in summer compared to winter. Especially, the phase of the circadian rhythm (measured using HR data [*HR Acrophase*]) was 67.9 minutes later in summer compared to winter. These findings mostly align with previous laboratory/survey-based studies. Prior sleep research has shown that people tend to sleep longer in winter than in other seasons, which may relate to the link between light exposure and melatonin production.[70-73] Previous studies have also found that individuals exhibited higher activity levels,[34, 36, 37] less fragmented rest-activity rhythms (lower intradaily variability),[74, 75] and delayed acrophase of the hormone secretion rhythm[76, 77] in summer compared to winter.

This study was performed on a relatively large cohort with a long study period, examining the effects of seasons and depression severity on circadian rhythms with a high temporal resolution. The use of three modalities from wearable data to estimate circadian rhythms offers a comprehensive investigation, with consistent patterns emerging across different modalities.

Notably, our results from wearable-measured circadian rhythms align with previous survey and laboratory studies, indicating the robustness of mobile technology in objectively monitoring behavior rhythms.

This study has several limitations. First, missing data may introduce bias, as our previous study found data compliance is associated with depression severity and other personal traits (e.g., age).[78] Second, our cohort with a history of depression and a majority of females, may limit the generalizability of our findings to general populations. Third, although we considered national lockdown as a covariate and performed a pre-COVID subset analysis, the effects of COVID-related restrictions varied across individuals and countries. Future validations are needed on post-COVID datasets.

## CONCLUSION

Our analysis of longitudinal wearable data from a large cohort underscores the significant seasonal impact on circadian rhythms and their associations with depression, suggesting seasonal effects should be considered in longitudinal mHealth studies. Also, wearable-measured circadian rhythms were found to be significantly associated with depression severity while controlling the seasonal effects, indicating they have the potential to be the digital biomarkers of depression. Our findings contribute valuable insights to our understanding of depression mechanism and pathology and provide the basis for future long-term health monitoring.

## ACKNOWLEDGEMENTS

The Remote Assessment of Disease and Relapse–Central Nervous System (RADARCNS) project has received funding from the Innovative Medicines Initiative (IMI) 2 Joint Undertaking under grant agreement No 115902. This Joint Undertaking receives support from the European Union's Horizon 2020 Research and Innovation Program and the European Federation of Pharmaceutical Industries and Associations (EFPIA). This communication reflects the views of the RADAR-CNS consortium and neither IMI nor the European Union and EFPIA are liable for any use that may be made of the information contained herein. The funding bodies have not been involved in the design of the study, the collection or analysis of data, or the interpretation of data. This study represents independent research partly funded by the National Institute for Health Research (NIHR) Maudsley Biomedical Research Centre at South London, and Maudsley NHS Foundation Trust and King's College London. The views expressed are those of the author(s) and not necessarily those of the NHS, the NIHR, or the Department of Health and Social Care. We thank all the members of the RADAR-CNS patient advisory board for their contribution to the device selection procedures, and their invaluable advice throughout the study protocol design. This research was reviewed by a team with experience of mental health problems and their careers, who have been specially trained to advise on research proposals and documentation through Feasibility and Acceptability Support Team for Researchers (FAST-R), a free, confidential service in England provided by the NIHR Maudsley Biomedical Research Centre via King's College London and South London and Maudsley NHS Foundation Trust. We thank all GLAD Study volunteers for their participation, and gratefully acknowledge the NIHR BioResource, NIHR BioResource centers, NHS Trusts and staff for their contribution. We also acknowledge NIHR BRC, King's College London, South London and Maudsley NHS Trust and King's Health Partners. We thank the NIHR, NHS Blood and Transplant, and Health Data Research UK as part of the Digital Innovation Hub Program. Participants in the CIBER site came from the following four clinical communities in Spain: Parc Sanitari Sant Joan de Déu Network services, Institut Català de la Salut, Institut Pere Mata, and Hospital Clínico San Carlos. Participant recruitment in Amsterdam was partially accomplished through Hersenonderzoek.nl (www.hersenonderzoek.nl), a Dutch online registry that facilitates participant recruitment for neuroscience studies. Hersenonderzoek.nl is funded by ZonMwMemorabel (project no 73305095003), a project in the context of the Dutch Deltaplan Dementie, Gieskes-Strijbis Foundation, the Alzheimer's Society in the Netherlands and Brain Foundation Netherlands.

**ETHICS APPROVALS**

Ethical approvals were obtained from the Camberwell St. Giles Research Ethics Committee (17/LO/1154) in the UK, the Fundacio Sant Joan de Deu Clinical Research Ethics Committee (CI: PIC-128-17) in Spain, and the Medische Ethische Toetsingscommissie VUmc (2018.012–NL63557.029.17) in the Netherlands.

**CONFLICT OF INTEREST STATEMENT**

S.V. is an employee of Janssen Research and Development LLC. V.A.N. was employed by Janssen Research and Development LLC during the duration of this study. P.A. was employed by the pharmaceutical company H. Lundbeck A/S during the duration of this study. D.C.M. has accepted honoraria and consulting fees from Apple Inc, Otsuka Pharmaceuticals, Pear Therapeutics, and the One Mind Foundation; has received royalties from Oxford Press; and has an ownership interest in Adaptive Health Inc. M.H. is the principal investigator of the Remote Assessment of Disease and Relapse–Central Nervous System project, a private public precompetitive consortium that receives funding from Janssen, UCB, Lundbeck, MSD, and Biogen. C.O. is supported by the UK Medical

**Figure 1. An example of a participant's processed HR, step, and sleep Fitbit data during the preceding 14 days of a PHQ-8 assessment (14-day PHQ-8 interval) collected via the RADAR-base platform.**

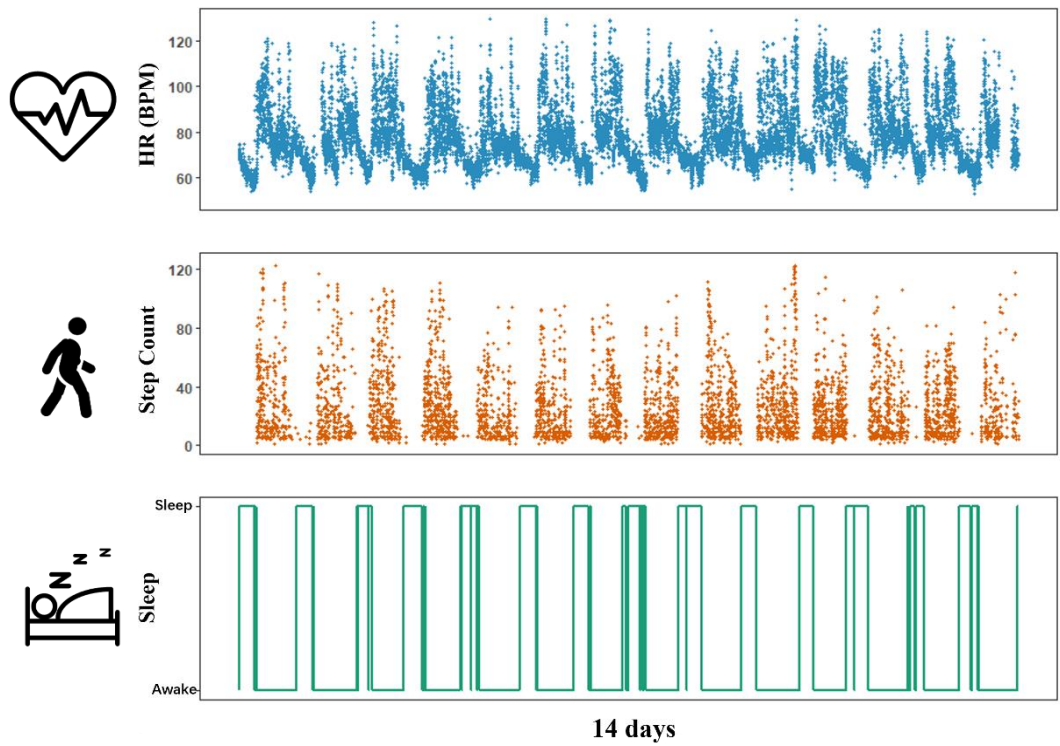

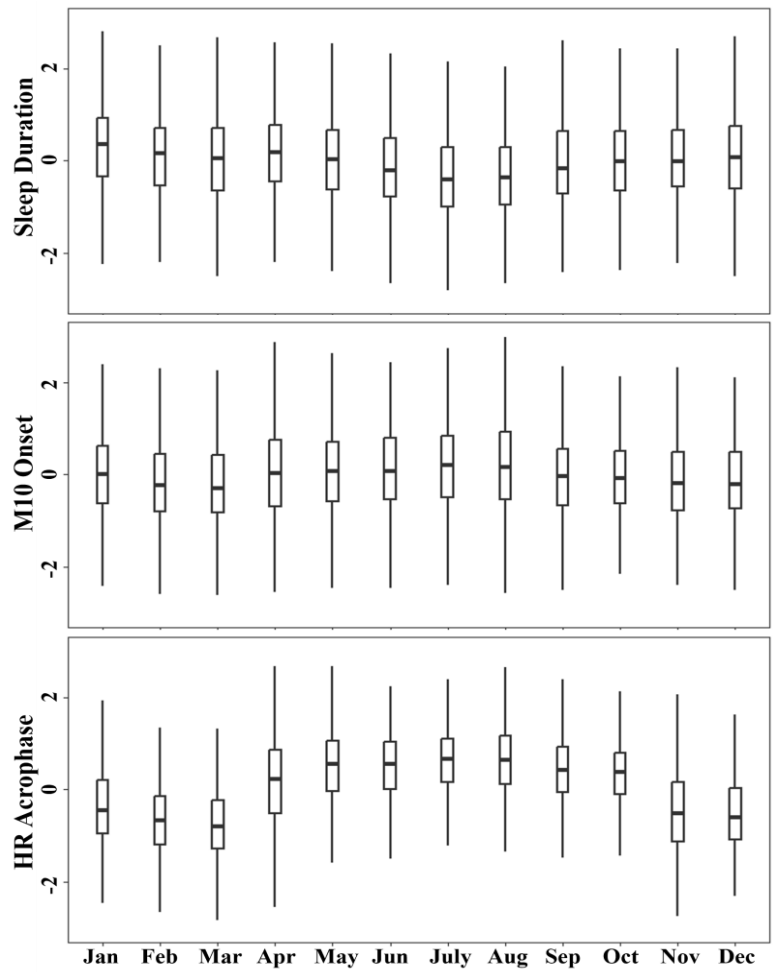

**Figure 2. Variations of wearable-measured circadian rhythm features across a year. For each participant, circadian rhythm features were normalized to reduce the individual differences.**